\documentclass{PoS}
 
 \title{Strongly and slightly flavored gauge theories}
 
 \ShortTitle{Strongly and slightly flavored gauge theories}
  
 \author{\speaker{Elisabetta Pallante}\\
        Centre for Theoretical Physics, University of Groningen, 9747 AG, Netherlands\\
        E-mail: \email{e.pallante@rug.nl}}
 
\abstract{We review some recent progress in our understanding of the phase diagram of non 
abelian gauge theories, by varying their flavor content -- fermion representations 
and the number of flavors. In particular, we explore the way
conformal symmetry can be restored before the loss of asymptotic freedom, and through 
a subtle interplay of perturbation theory, chiral dynamics and confining forces. 
It is with the combination of numerical lattice studies and theoretical insights into gauge 
theories with and without supersymmetry that we may successfully attempt to clarify the 
missing pieces of this puzzle.   }
 
 \FullConference{The XXVII International Symposium on Lattice Field 
Theory\\
                 July 26-31, 2009\\
                 Peking University, Beijing, China}
 
\begin{document}
 
 \section{The phase diagram of gauge theories and conformal symmetry}

This review explores non abelian gauge theories beyond what is the best 
known, yet challenging example among others, quantum chromodynamics (QCD).
$SU(N)$ Yang Mills theory, with $N$ gauge degrees of freedom -- the colors -- 
and no matter content, is asymptotically free and confines. 
Fermionic matter, added in various representations of $SU(N)$, effectively screens the 
strength of gauge interactions and eventually modifies the 
fundamental properties of the theory. First, confinement will disappear and chiral symmetry restores, then asymptotic freedom will be lost.
This change of properties can be attributed to the emergence of fixed points of the 
renormalization group flow of the theory -- in the infrared or the ultraviolet -- 
accompanied by the occurrence of phase transitions in parameter space.
Uncovering such change of properties thus consists in building a detailed map of 
the phase diagram of these theories in terms of their relevant parameters: 
these are the temperature $T$, the number of fermionic 
flavors $N_f$ and the gauge coupling constant $g$.

Conformal symmetry plays as protagonist in the zero-temperature plane. This 
symmetry is easily lost in four dimensional field theories, but can eventually be restored 
by varying their matter content. The presence of a conformal fixed point with anomalous 
dimensions can be a relevant ingredient for models of electroweak interactions at energies 
beyond the Standard Model. The running of gauge couplings is slowed down in a nearly 
conformal region of the parameter space, and it is this feature that has inspired
 new versions of technicolor models, such as tumbling \cite{Tumbling} and walking 
technicolor \cite{Walking}.
On the other hand, understanding the way conformal symmetry is recovered, the relevance of 
the AdS/CFT correspondence for this mechanism, and the 
differences and similarities of theories with and without supersymmetry in this context,  
has a theoretical appeal in its own.
As may be expected and based on our knowledge of QCD, chiral dynamics plays a major role in 
the game, and eventually renders a 
perturbative study insufficient. Then, a lattice formulation of gauge theories can in 
principle provide a non perturbative answer to the problem.

In the search for conformality, a distinction can be made between $SU(N)$ gauge 
theories with fermions in the fundamental representation and those with  
fermions in higher dimensional representations, such as the adjoint or two-index symmetric. 
Properties of the first ones need many flavors to significantly change and recover 
conformality; I thus call these theories strongly flavored.
Properties of the latter change when just a few flavors are added, and I call them 
slightly flavored theories.
 
The rest of this review is organized as follows. Section \ref{sec:strongly} is devoted to 
strongly flavored theories, meaning the case of a QCD-like theory with an 
enlarged flavor content. I will review some basic knowledge of the perturbative $\beta$ 
function, formulate a few paradigms that might serve as guidance for lattice studies, and   
summarize recent lattice results and analytical conjectures.
Section \ref{sec:slightly} treats the case of $SU(N)$ gauge theories with fermions in 
higher dimensional representations. Possible scenarios for the zero-temperature phase 
diagram are discussed and recent lattice studies reviewed.
I conclude in Section \ref{sec:conc} with discussing future prospects, and 
connections to the physics at the Large Hadron Collider (LHC).
 
\section{Strongly flavored theories: $SU(3)$ with fundamental fermions}
\label{sec:strongly}

\begin{figure}
\begin{center}
\includegraphics[width=.5\textwidth]{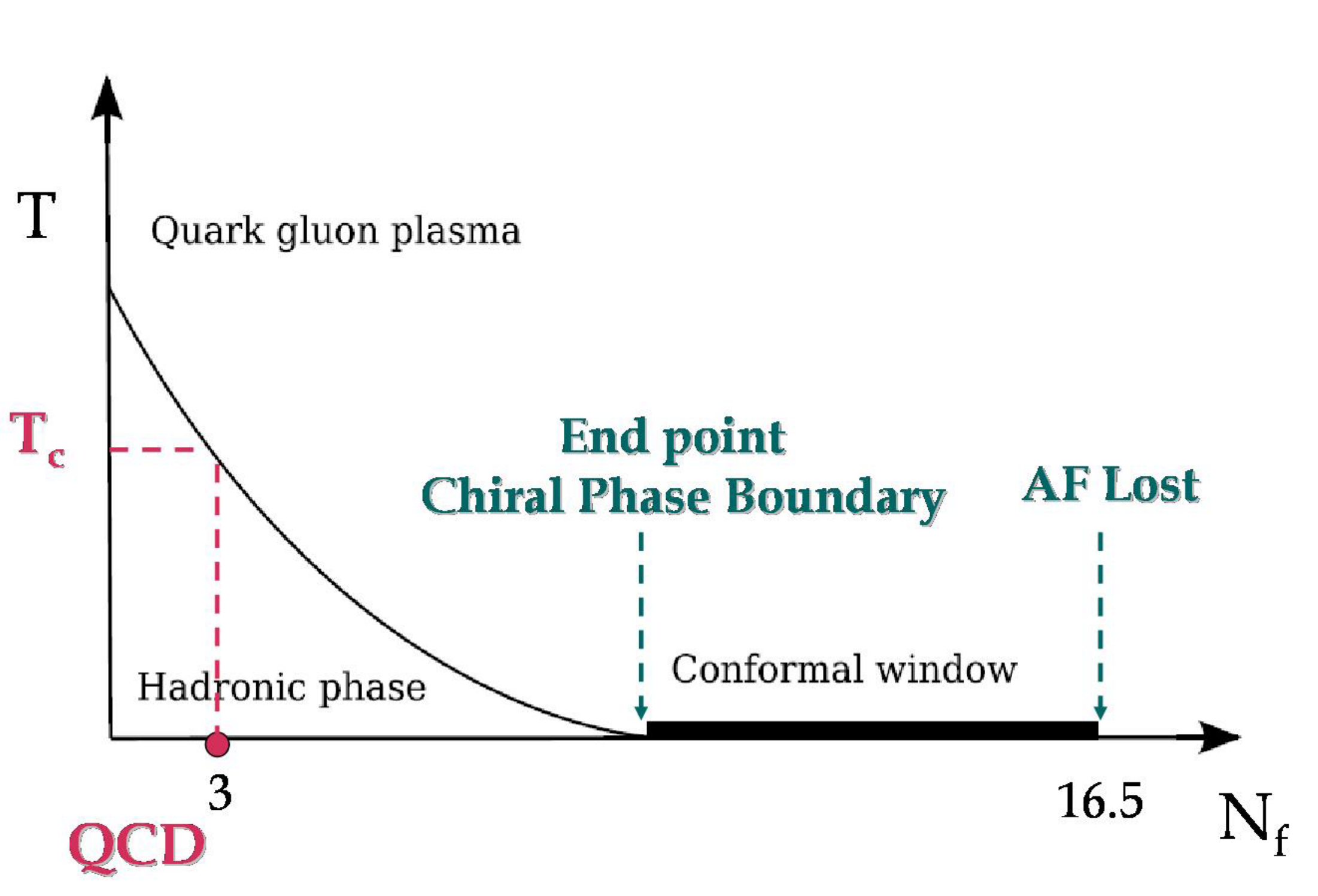}
\caption{A qualitative view of the phase diagram of a QCD-like gauge theory with varying 
flavor content, in the $T, N_f$ plane. }
\label{fig:PhaseD_TNf} 
\end{center}
\end{figure}

We start with QCD with massless fermions and increase its flavor content.
The first instructive step is to project the three-dimensional phase diagram onto the plane 
temperature $T$ and number of flavors $N_f$. Fig.~\ref{fig:PhaseD_TNf} provides a 
qualitative summary of what we know and what we reasonably expect to be true. 
We know that QCD with three light flavors, at zero temperature, is in the hadronic 
phase: chiral symmetry is broken and the theory confined. At a critical temperature 
$T_c\sim 120$~MeV, predicted by lattice QCD calculations, a phase transition 
occurs to quark-gluon plasma (QGP), where chiral symmetry is restored and the theory 
deconfined. Properties of QGP can be inferred from conformal symmetry to a good 
approximation.
On the right hand side of Fig.~\ref{fig:PhaseD_TNf}, at zero temperature and larger $N_f$, 
we know that asymptotic freedom is lost at 
$N_f = 16 {1\over 2}$~\cite{AsymFree}. 
In the intermediate region of this phase diagram,
we envisage the existence of a continuous line of finite temperature phase transitions 
between a chirally broken (low-T) and a chirally symmetric (high-T) phase.
The end point of this chiral phase boundary signals the opening of the 
{\em conformal window} at zero temperature\footnote{For an analytical study of the scaling 
of chiral observables near the end point of the chiral phase boundary see 
\cite{BraunGies}.}. 
The structure itself of the phase diagram 
suggests that the appearance of a conformal window is likely to result from the   
interplay between perturbation theory and genuinely non-perturbative chiral dynamics.
We are thus challenged with a series of interesting questions where perturbative and non 
perturbative dynamics are interwound, and lattice field theory 
can provide useful answers. In particular, where is the end point of the chiral phase 
boundary located, what is the extent of the conformal window, how the zero temperature 
conformal region at large $N_f$ is connected to the high temperature, small $N_f$ 
quark-gluon plasma? The nature of chiral and deconfining phase transitions at 
various points of the phase diagram is also an historically challenging problem. 

\subsection{The perturbative running of the gauge coupling}

The first step to understand the emergence, or loss, of conformality is to investigate the 
perturbative renormalization group (RG) flow of the theory in parameter space, in this case 
the running of the gauge coupling $g$ in the weak coupling regime. 
It suffices to consider the Callan-Symansik beta function to two loops, calculated 
in~\cite{Caswell}:
\begin{eqnarray}
\label{eq:beta}
&&\beta (g) = -b_0 \frac{g^3}{16\pi^2}\, -\, b_1 \frac{g^5}{(16\pi^2)^2} \, + O(g^7)
\nonumber\\
&&b_0 = \frac{11}{3}C_2(G)-\frac{4}{3}T(R)N_f\nonumber\\
&&b_1= \frac{34}{3}C_2(G)^2-\frac{20}{3}C_2(G)T(R)N_f-4C_2(R)T(R)N_f\, ,
\end{eqnarray}
for $N_f$ massless Dirac fermions in the representation $R$ of the compact Lie gauge group 
$G$. 
$C_2(G)$ and $C_2(R)$ are the quadratic Casimir operators of the adjoint and fermion 
representations, respectively, and $T(R)$ is the trace of $R$.
\begin{figure}
\begin{center}
\includegraphics[width=.5\textwidth]{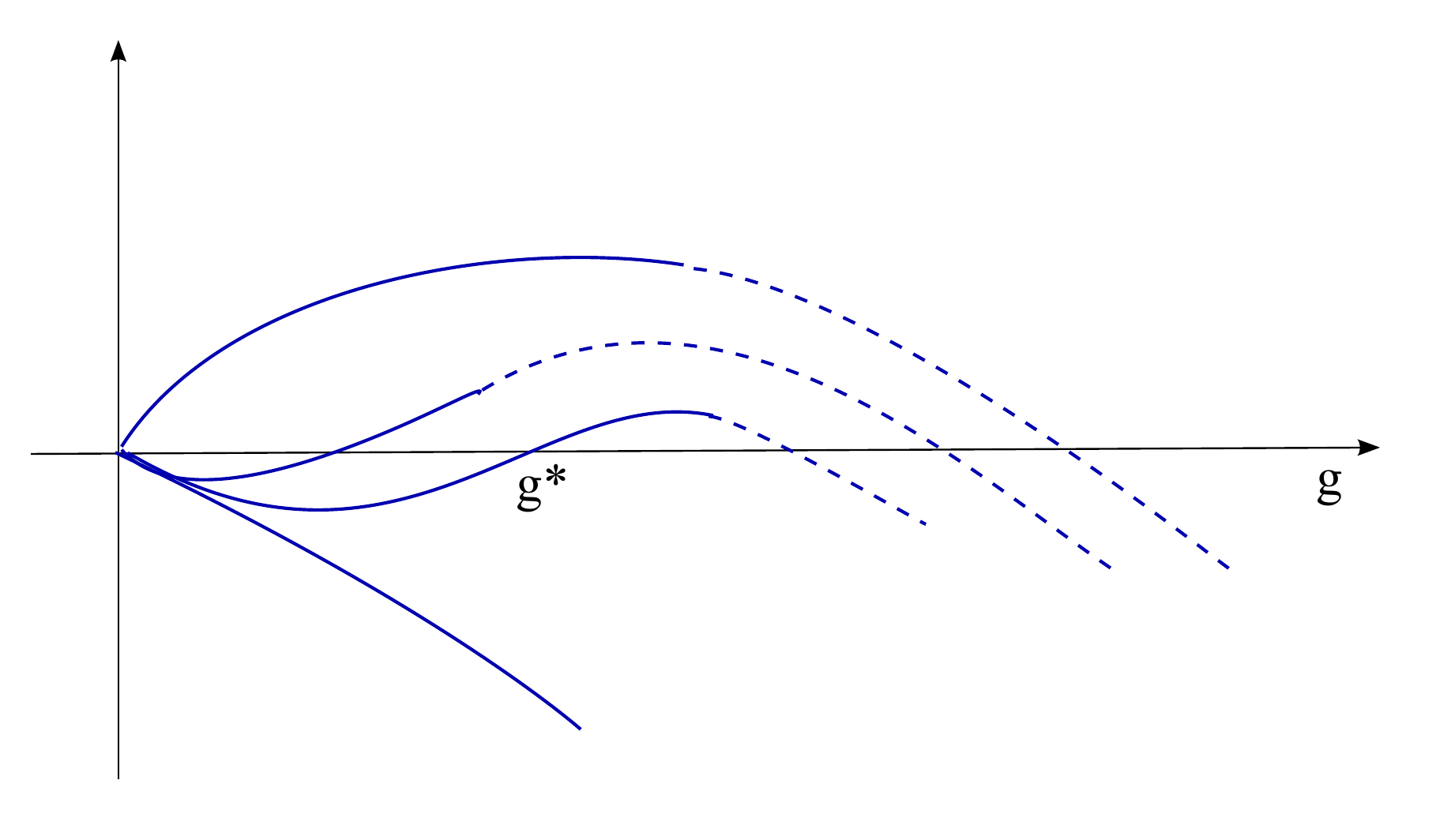}
\caption{The behavior of the two-loop beta function $\beta (g)$ as a function of the 
coupling $g$ for $SU(3)$, and increasing (bottom to top) the number of massless flavors 
$N_f$ in the fundamental representation. 
For $N_f<8.05$ the beta function stays negative. For $8.05<N_f<16.5$ the $\beta$ function 
develops a zero (IRFP) at a non zero coupling $g^*$. For $N_f>16.5$, the $\beta$ function 
becomes positive, implying that $g=0$ is not anymore an UV stable fixed point and asymptotic 
freedom is lost. The beta function might develop an additional zero (UVFP) at strong 
coupling (dashed lines), a property to be investigated in future studies. }
\label{Betafunction}
\end{center}
\end{figure}
The coefficients $b_0$ and $b_1$ are universal, meaning they are renormalization scheme
independent and their flavor dependence makes evident how the matter content affects, 
depending on the representation, the sign and zeros of the $\beta$ function, modifying the 
RG flow of the theory. 
For $G=SU(3)$ and $N_f$ Dirac fermions in the fundamental representation 
the first two coefficients read $b_0=11-{2}N_f/3$ and $b_1=102-{38}N_f/3$ -- 
see Table \ref{tab:R}.
\begin{table}[b]
\begin{center}
\begin{tabular}{|c|c|c||c|c|c|}
\hline
$R$ & $C_2(R)$  & $T(R)$ & $R$ & $C_2(R)$  & $T(R)$ \\
\hline
Fund & $\frac{N^2-1}{2N}$   & ${1}/{2}$ 
& 2S & $\frac{(N-1)(N+2)}{N}$   & $\frac{N+2}{2}$  \\
Adj(G) & $N$   & $N$ & 2A & $\frac{(N+1)(N-2)}{N}$   &   $\frac{N-2}{2}$  \\
\hline
\end{tabular}
\end{center}
\caption{The quadratic Casimir operator $C_2(R)$ and the trace $T(R)$ for various irreducible representations $R$ of $SU(N)$: fundamental, adjoint, two-index symmetric (2S) and two-index antisymmetric (2A). }
\label{tab:R}
\end{table}
For $N_f^{AF}=11C_2(G)/4T(R)=16.5$, $b_0$ becomes negative and asymptotic freedom is lost, 
see Fig.~\ref{Betafunction}. 
It should be noted that, if $b_0=0$, the first non zero contribution to the beta function is
\begin{equation}
\beta (g)=  \frac{g^5}{(16\pi^2)^2}\left ( 7C_2(G)^2+11C_2(G)C_2(R) \right)\, , 
\end{equation}
which is positive for any group $G$. Thus, at least perturbatively, we should not expect 
the occurrence of a zero at finite coupling, signaling an ultraviolet fixed point (UVFP) 
in the non asymptotically free theory. 
Its non perturbative existence in the continuum is however not excluded, it was already 
discussed in early works \cite{BanksZaks}, recently reconsidered \cite{Kaplan} and 
treated in Section~\ref{sec:conj}.
The second coefficient $b_1$ changes its sign at $N_f^*<N_f^{AF}$, given by 
\begin{equation}
N_f^*=  \frac{34}{3}\frac{C_2(G)^2}{T(R)\left ( \frac{20}{3}C_2(G)+4C_2(R)\right )}
\end{equation}
and $N_f^*\simeq 8.05$ for $SU(3)$ with fundamental fermions. 
It implies the emergence of a simple zero of the two-loop beta function at finite 
coupling, i.e. a stable infrared fixed point (IRFP) where the theory is conformally 
invariant with anomalous dimensions and the coupling approaches its fixed-point 
value with a power law.

A perturbative expansion in the small parameter $N_f^{AF}-N_f$ was 
suggested in~\cite{BanksZaks}. Such an expansion implicitly assumes that we are 
allowed to consider non integer $N_f$: we can indeed continue $N_f$ to non integer values 
since it appears analytically in the path integral of the theory. 
For $N_f\lesssim N_f^{AF}=16.5$ perturbation theory thus implies the 
existence of an entire family of asymptotically free theories with a finite zero of the 
beta function and conformal behavior. Approaching $N_f=8$ the task of establishing the 
emergence of conformality becomes increasingly non perturbative. 

\subsection{The theory beyond perturbation theory}

\begin{figure}
\begin{center}
\includegraphics[width=.65\textwidth]{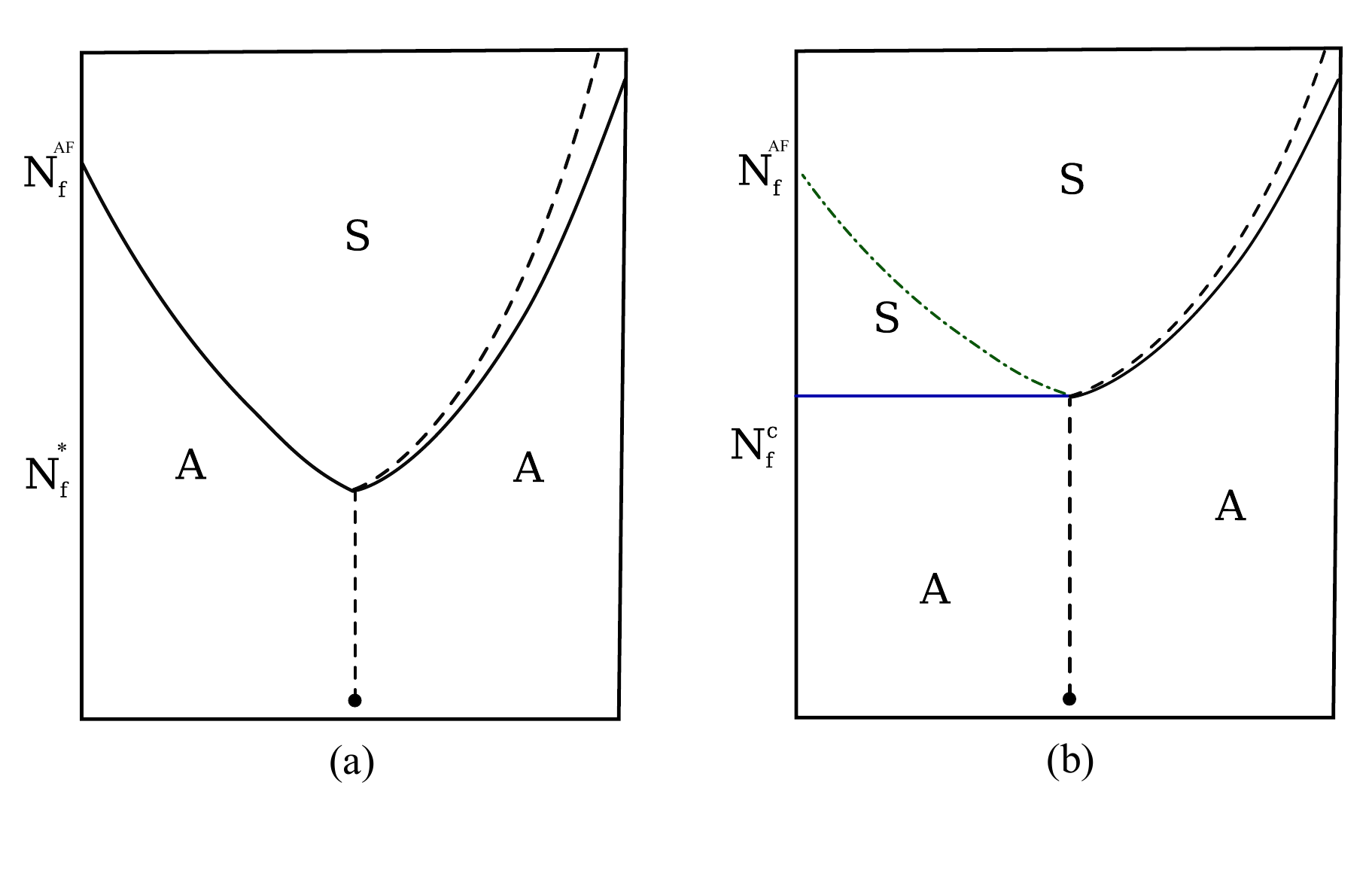}
\caption{(a) The Banks-Zaks scenario for conformal symmetry restoration~\cite{BanksZaks}
by varying the gauge coupling $g$ and flavors $N_f$. S and A refer to chirally symmetric and asymmetric, respectively. Dashed lines are chiral transitions and solid lines are confinement 
transitions. They coincide on the left branch, location of the IRFP. (b) The modified 
scenario by \cite{Miransky_conformal_1997}. The 
horizontal (blue) line is the conformal phase transition (CPT) that opens up the conformal window. The dot-dashed (green) line is the location of the IRFPs.
}
\label{fig:Miransky_plot}
\end{center}
\end{figure}

Physics considerations, together with early results from explorative lattice studies,
inspired the scenario proposed by Banks and Zaks~\cite{BanksZaks} and reported in 
Fig.~\ref{fig:Miransky_plot}(a), for an $SU(N)$ gauge theory with fundamental fermions. 
Projecting the phase diagram onto the zero-temperature plane 
$(g,N_f)$ -- where $g$ is the bare lattice coupling and the phase diagram can eventually be 
mapped into the continuum --, the following features were envisaged. 
The line to the left in Fig.~\ref{fig:Miransky_plot}(a) is the location of the infrared 
fixed points associated with the zero 
of the perturbative beta function, where larger flavors correspond to weaker couplings. 
The line starts at $N_f^*$ and ends at $N_f^{AF}$. 
Chiral symmetry should be exact on the right of this line, and the theory deconfined. 
The authors of \cite{BanksZaks} thus inferred that a deconfinement first order 
phase transition should accompany the emergence of the IRFP, together with a chiral 
transition from a broken to a symmetric phase. If driven by instabilities, 
the latter could also presumably be a first order transition. This picture, however, would 
lead to the counterintuitive conclusion that, for a given number of flavors, a phase 
transition from a chirally broken to a symmetric phase is taking place while 
moving towards stronger couplings.

On the right side of Fig.~\ref{fig:Miransky_plot}(a), a second order confining transition 
 corresponding to the emergence of an UVFP -- an additional zero of the continuum 
beta function -- was also envisaged. This line extends to 
non asymptotically free theories with $N_f>N_f^{AF}$, and carries flavor dependence, since
 larger $N_f$ produce more effective screening of confining forces, pushing the transition 
to stronger couplings. A chiral phase transition should also occur: 
fermions in the fundamental representation do feel confining forces, thus the chiral 
condensate cannot be zero as long as the transition to the massless 
gluon phase (deconfined phase) has not occurred. Thus the chiral line can only be to the 
left of the confinement line. Chiral symmetry breaking ($\chi$SB) and confinement are 
entangled for fundamental fermions, and lattice studies to date favor the conclusion that 
the two transitions actually coincide.

It is also true that the lattice regularized theory at zero temperature 
will undergo a transition to a chirally broken phase at 
sufficiently strong coupling, usually referred to as a bulk transition. 
It is a task for the lattice to distinguish between a scenario with a strong coupling 
UVFP in the continuum theory, and thus accompanied by second order phase transitions, and 
a scenario with a lattice bulk transition to a phase with no continuum limit
-- see also Section~\ref{sec:conj}.
The two lines in Fig.~\ref{fig:Miransky_plot}(a) plausibly coalesce at one point $(N_f^*,g^*)$, below which the theory always 
confines and is chirally broken.
A first order bulk transition, a feature dependent on the lattice action, between two 
chirally broken phases can occur at lower number of flavors, and should be absent in the 
pure gauge theory. 

It is the study of the dynamics of chiral symmetry that brought to the discovery of the 
{\em conformal window} in \cite{Appelquist_1996, Miransky_conformal_1997}. 
In~\cite{ Miransky_conformal_1997} a line is added to the phase diagram of 
Fig.~\ref{fig:Miransky_plot}(a), providing the new scenario in
Fig.~\ref{fig:Miransky_plot}(b), and avoiding the occurrence of a chiral restoration 
transition at the IRFP. 
The key ingredient is the observation that there exists a critical coupling,
solution of the Schwinger-Dyson gap equation, above which chiral symmetry must be 
broken.
This translates into a critical number of flavors $N_f^c$ -- the horizontal line of 
Fig.~\ref{fig:Miransky_plot}(b) -- at which a transition from a chirally symmetric 
many-flavor phase to a chirally broken phase occurs. 
Such a zero-temperature phase transition \cite{Appelquist_1996, Miransky_conformal_1997} 
is not of second order, nor first order, and named 
{\em conformal phase transition} (CPT): it is continuous in the chiral order parameter with 
an abrupt change in the spectrum of the theory. 
A CPT might also enter in the dynamics of strongly coupled condensed matter systems, such 
as a non-Fermi liquid, which might be relevant for high-temperature superconductivity. 

The CPT at $N_f^c$ signals the opening of the conformal window, the interval 
$N_f^c<N_f<N_f^{AF}$ 
where theories develop a conformal IRFP with anomalous dimensions, they do not confine and 
chiral symmetry is exact. The green line in Fig.~\ref{fig:Miransky_plot}(b) is the 
location of the IRFPs where, differently from Fig.~\ref{fig:Miransky_plot}(a), no phase 
transition occurs. It separates two chirally symmetric and deconfined theories which differ 
in their short 
distance behavior: an asymptotically (ultraviolet) free theory to the weak coupling side, 
and a coulomb theory (infrared free) to the strong coupling side. 
Besides separating two symmetric phases, the phase diagram of 
Fig.~\ref{fig:Miransky_plot}(b) differs from Fig.~\ref{fig:Miransky_plot}(a) 
because the lower end point of the new phase occurs at $N_f^c$ and not at $N_f^*$, the value 
predicted by the perturbative beta function; 
chiral dynamics has been incorporated, and for $N_f<N_f^c$ the infrared stable fixed point 
is washed out by the generation of a dynamical fermion mass.

The first quantitative prediction for the lower end-point of the conformal window was 
obtained \cite{Appelquist_ladder} by solving the 
Schwinger-Dyson gap equation in the ladder approximation, and demanding
that the conformal window should close when the IRFP coupling $\alpha (N_f)$ -- 
predicted by the perturbative beta function -- reaches the critical value
at which a mass gap arises; this happens when the anomalous dimension of the fermion mass
operator $\bar{\psi}\psi$ is $\gamma =1$.
$N_f^c$ can thus be determined at a given order in the combined 
perturbative expansion of the beta 
function and the anomalous dimension $\gamma$. 
 
To leading order and for $SU(3)$ with fundamental fermions $N_f^c\simeq 11.9$\footnote{ 
An enhancement of the chiral condensate w.r.t. 
the ordinary QCD phase is also predicted in the broken phase and in the vicinity of the 
critical point $N_f\lesssim N_f^c$ \cite{Appelquist_ladder}.}.  
As the authors of \cite{Appelquist_ladder} interestingly observed,  
the ladder approximation becomes gradually more accurate 
for higher dimensional representations, and suggest that the estimate for fundamental 
fermions is good to about $20\%$. The question remains, if the mechanism advocated above 
and entirely driven by chiral dynamics is actually providing the complete picture.   
The uncertainties involved in the analytical predictions, and the subtle interplay of chiral 
dynamics and confinement invite for a lattice enterprise.

\subsection{Lattice studies}

Two complementary strategies can pave the way for lattice studies, and have in fact been 
implemented in recent works.
One can reconstruct the theory renormalization group flow on the lattice by 
means of the discretised beta function, followed by the recovery of the continuum limit.
This can be done with the Schroedinger functional and step scaling function technique, as 
in~\cite{Appelquist_RG}.
Similarly, a Monte Carlo RG can be formulated on the lattice~\cite{Hasenfratz}.
The aim of this type of studies is the mapping of the zero's of the  beta function; 
the emergence of conformality -- alias the existence of a stable IRFP -- is probed by 
the flattening of 
the continuuum running coupling, reached from the far ultraviolet and the far infrared.

The other strategy, complementary to the previous, is inspired by the physics 
of phase transitions, and is driven by the general scope of uncovering the complete 
phase diagram of gauge theories. This approach has been recently adopted 
in~\cite{Deuzeman_12}, 
and an early study can be found in~\cite{Heller}. 
Chiral symmetry and its breaking pattern is also monitored by the distribution of the 
eigenvalues of the Dirac operator\footnote{The assumption of preserved 
maximal flavor symmetry implies a unique breaking 
pattern for all irreducible representations\cite{Peskin}. In particular 
 $SU(N_f)\times SU(N_f)\to SU(N_f)_V$ for complex,  
$SU(2N_f)\to O(2N_f)$ for real, and  $SU(2N_f)\to Sp(2N_f)$ for pseudoreal 
representations. Staggered fermions on coarse lattices show an inverted pattern for real 
and pseudoreal fermions~\cite{Svetitsky}. };  this strategy has been investigated in 
\cite{Kuti_12}.

\begin{figure}
\begin{center}
\includegraphics[width=.5\textwidth]{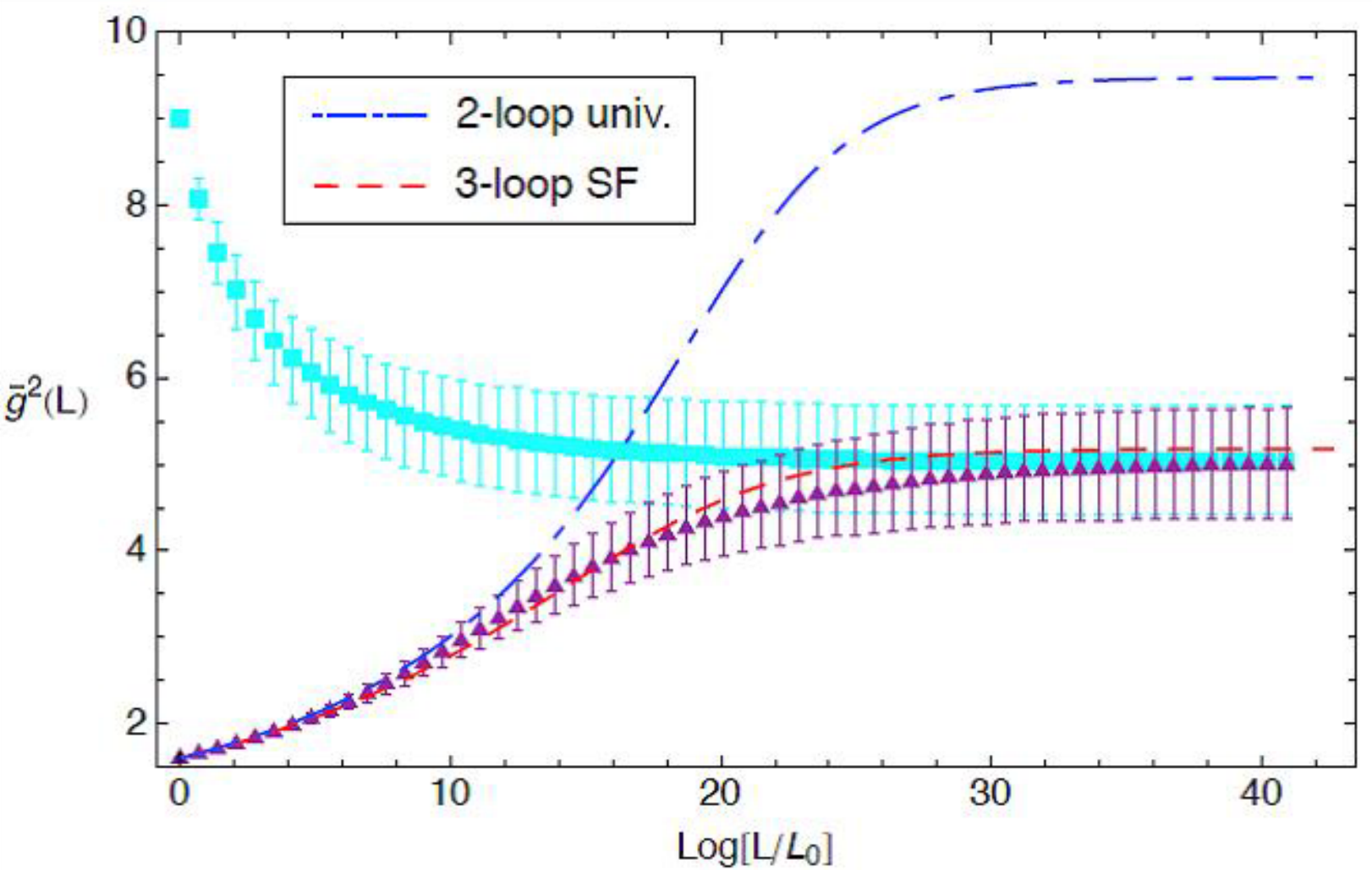}
\caption{The running gauge coupling for $SU(3)$ with twelve flavors of fermions in the 
fundamental representation from \cite{Appelquist_RG}. }
\label{fig:Fleming_RG}
\end{center}
\end{figure}

Recent studies have established that $N_f=8$ lies within the hadronic phase of 
QCD~\cite{Deuzeman_8}, resolving a long time contention.
Delicate task remains the one of locating the lower-end point of the conformal 
window. Analytical studies favor a value around $N_f=12$ \cite{Appelquist_ladder,BraunGies}, 
reason for recent lattice investigations with twelve 
flavors~\cite{Appelquist_RG,Deuzeman_12,Hasenfratz,Kuti_12,Mawhinney_12}. The result 
reported by \cite{Appelquist_RG} and summarized in Fig.~\ref{fig:Fleming_RG} beautifully 
shows a flattening of the running coupling from the far infrared (from above) and the far 
ultraviolet (from below), thus suggesting that twelve flavors are in the conformal window. 
The same conclusion is reached by the authors of~\cite{Deuzeman_12}, based on the 
physics of phase transitions and a strategy that uses as a paradigm the scenario of 
Fig.~\ref{fig:Miransky_plot}(b). The strategy allows to explore the theory in different 
regimes and regions of the phase diagram, and infer the existence of the IRFP from outside 
its basin of attraction. Three ingredients are used: moving from strong to weak coupling one 
encounters i) a bulk transition in the chiral order parameter, ii) a high-statistics, 
infinite volume chiral extrapolation 
of the condensate strongly favors exact chiral symmetry on the weak coupling side of the 
transition and iii) the mass specrum is compatible with a positive beta function, signature 
of a Coulomb phase. This result agrees with Fig.~\ref{fig:Miransky_plot}(b), while the 
exclusion of Fig.~\ref{fig:Miransky_plot}(a) further needs the confirmation that no 
$\chi$SB transition is encountered at weaker couplings. 
Other numerical studies \cite{Hasenfratz,Kuti_12,Mawhinney_12} have challenged the 
conclusion of~\cite{Appelquist_RG, Deuzeman_12}. This is hardly surprising, given that 
$N_f=12$ is likely to be close to $N_f^c$, making a numerical study particularly delicate.
In this context, I would like to observe that a comparison of the light spectrum and decay 
constants of the 
theory with the form predicted by chiral perturbation theory -- at large or small volumes --
might lead to incomplete or misleading intepretations and needs to be corroborated by 
additional evidences such as phase transitions and high precision chiral extrapolations.   

Alternatively, one can isolate optimal observables that are directly sensitive to the 
presence of an infrared fixed point and work within its basin of attraction; 
recalling from Fig.~\ref{fig:Miransky_plot}(b) that theories on the 
sides of the infrared fixed point differ in their ultraviolet behavior -- the presence or 
absence of asymptotic freedom -- one should expect that observables sensitive to the far 
ultraviolet will show a change in behavior across the fixed point. This might be the case 
for the plaquette, or particular combinations of its spatial and temporal components --
entangled to the trace anomaly \cite{Progress}. 
The inclusion of additional scales, mass and temperature, can be used to measure the 
anomalous dimensions of the relevant operators in the surroundings of the would be 
conformal fixed point. A power-law scaling of the chiral condensate 
$\langle\bar{\psi}\psi\rangle \sim A\,m^\delta$, determined by its 
anomalouos dimensions, should be expected in the vicinity of the infrared fixed point 
\cite{Sannino_AD}; this behavior is in fact observed in \cite{Deuzeman_12}.

Finally, a direct numerical study of the realization of a conformal phase transition by 
varying the number of flavors would be welcome; it is in principle feasible, provided the 
lattice fermion action can be extrapolated to the correct continuum limit for any value of 
the flavor number.

\subsection{Theory and conjectures: recent developments}
\label{sec:conj}

A closer comparison with supersymmetric gauge theories, the application of the AdS/CFT 
correspondence, and attempts to better account for the confining dynamics are all at the 
base of most recent developments in the theoretical description of conformality in non
 abelian non supersymmetric gauge theories. 
The phase diagram in Fig.~\ref{fig:Miransky_plot}(b), with the prediction of a conformal 
window, resembles what happens in 
supersymmetric theories; there, electromagnetic duality allows to establish the existence 
of a conformal window over the interval $3/2N_c<N_f<3N_c$ for supersymmetric 
QCD~\cite{Seiberg}, while the presence of supersymmetry provides an exact 
beta function also in the presence of matter multiplets~\cite{NSVZ} -- 
the NSVZ beta function. Crucial ingredients, 
consequences of supersymmetry, are the cancellation of non zero modes and non 
renormalization theorems, which guarantee for the anomalous dimensions 
$\gamma_{gluino} =\gamma_{gluon} = \beta (g) /\alpha (g)$.
Without supersymmetry, an NSVZ-like beta function is in general not expected, and 
one has to rely on truncations of the perturbative expansion, on 
particular symplifying limits such as the large-$N$ limit, or 
conjectures. One exception is the large-$N$ limit of $SU(N)$ with one Dirac flavor in the 
two-index antisymmetric or symmetric representation \cite{Armoni}, due to planar 
equivalence with supersymmetric Yang-Mills.
In all other examples, it is instructive to look for analogies and differences with the 
supersymmetric case. 
    
The beta function of Yang Mills theories in the large-$N$ limit recently derived 
in~\cite{Bochicchio}, is a potentially important result. The absence of supersymmetry -- 
and the presence of non zero modes -- is responsible for the appearance of an anomalous 
dimension term in the running of the canonical coupling $g_c=\sqrt{N}g$ 
and a renormalization scheme dependence, contrary to supersymmetric Yang-Mills.
It would be interesting to extend the result of~\cite{Bochicchio} to include matter fields, 
for example in the Veneziano limit in which $N_f\to \infty$ and $N_c\to\infty$ with 
$N_f/N_c$ finite\footnote{A possible generalization that reproduces the two-loop beta 
function in the Veneziano limit can be
\begin{equation}
\label{eq:beta_mine}
\beta (g_c) = \frac{-\beta_0^\infty g_c^3 +\frac{\beta_j}{4}g_c^3\left ( 
\frac{\partial\log Z}{\partial\log\Lambda} +c_F \frac{g_c^2}{16\pi^2} 
\right ) + c_F \frac{g_c^3}{16\pi^2}(1+\gamma (g_c^2)/2) }
{1-\beta_jg_c^2  }\, ,
\end{equation}
where we assumed the presence of matter fields only in the numerator in this limit, 
$c_F = 4 T(R)N_f/(3N)$, while $\beta_j$ and $\partial\log Z/\partial\log\Lambda$ are 
derived in~\cite{Bochicchio}. }.
On the other hand, the supersymmetry inspired beta function for $SU(N)$ gauge theories 
conjectured in~\cite{Sannino_Susy} 
and used to estimate a bound for the lower end of the conformal window, does not account
for the presence of an anomalous dimension term arising from the lack of supersymmetry. 
The presence of such a term would possibly modify the constraint
on the lower end of the conformal window predicted in~\cite{Sannino_Susy} 
and would force an analysis order by  order in the gauge coupling. 
\begin{figure}
\begin{center}
\includegraphics[width=.5\textwidth]{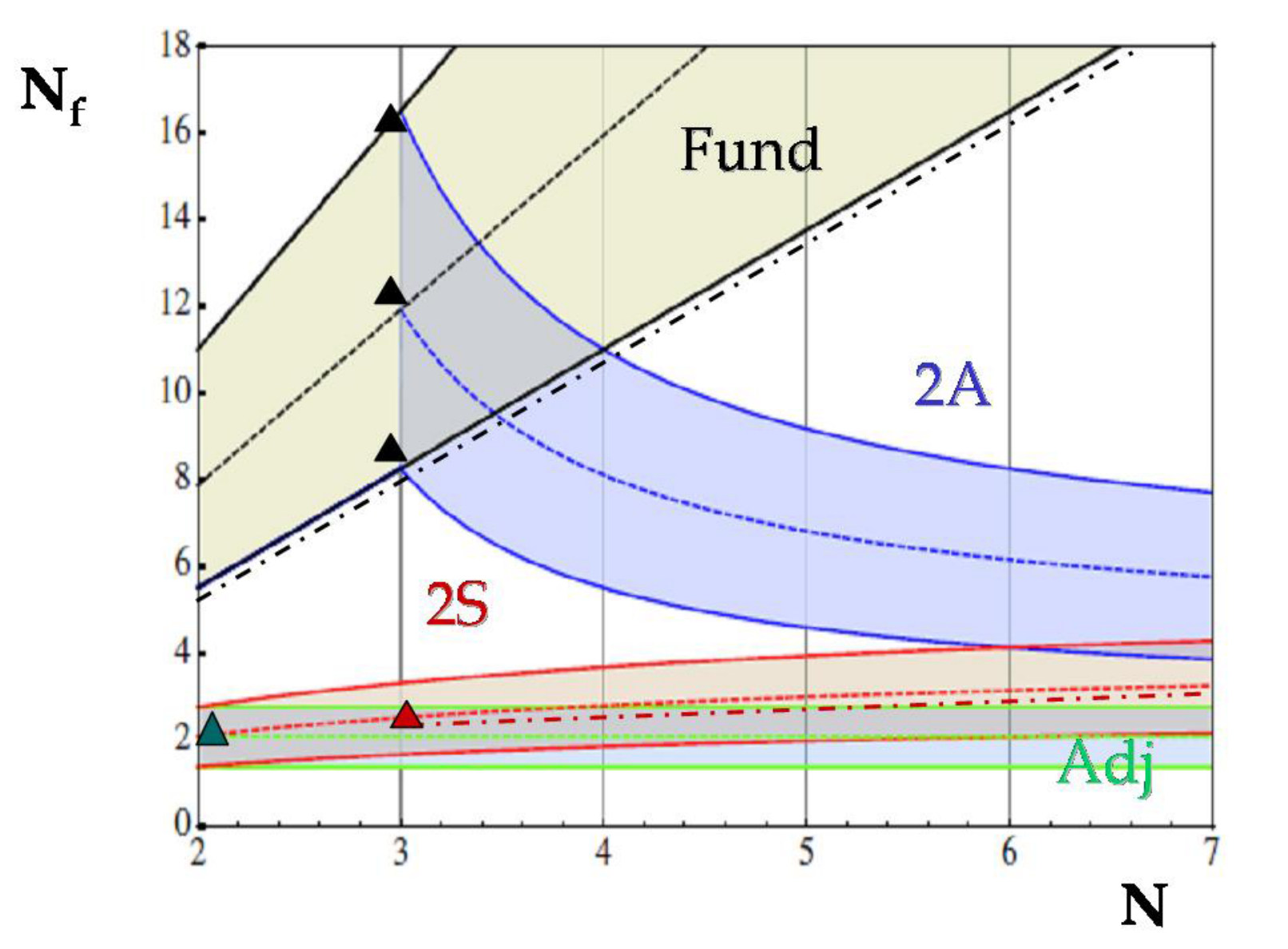}
\caption{Plot of the predicted conformal windows, adapted from \cite{Sannino_Susy}. 
For each representation, fundamental (Fund), two-index asymmetric (2A), two-index symmetric 
(2S) and adjoint (Adj), the lowest curve is the lower bound predicted by a supersymmetry 
inspired conjecture \cite{Sannino_Susy}, the middle curve (dashed) 
is the lower bound from the gap equation in the ladder approximation \cite{Appelquist_ladder}
 and the dot-dashed line is the prediction from deformation 
theory in \cite{Poppitz}. A more recent analysis \cite{Poppitz_2} moves the dot-dashed line
for fundamental fermions to $\sim 4N$, close to the ladder estimate. 
Triangles are points simulated on the lattice.}
\label{fig:ConfPlot}
\end{center}
\end{figure}

The upper part of Fig.~\ref{fig:ConfPlot}
summarizes the bounds on the conformal window obtained for fundamental 
fermions by using i) the ladder solution to the gap equation
\cite{Appelquist_ladder}, ii) deformation theory (DT) \cite{Poppitz}, and iii) a  
supersymmetry inspired conjecture for the beta function \cite{Sannino_Susy} to 
which the unitary bound on the dimension of $\bar{\psi}\psi$, 
$D_{\bar{\psi}\psi} =3-\gamma \geq 1$, is imposed.
The latter and the deformation theory analysis of \cite{Poppitz}
predict a lower bound around $N_f \sim 8$, similar to the two-loop beta function. 
The solution to the gap equation is more restrictive and provides a lower bound 
$N_f \sim 12$ that captures the bulk of chiral dynamics. 
No surprise that the unitarity bound is loosely constraining, since it corresponds to an 
anomalous dimension $\gamma =2$ of the fermion
mass operator, while the ladder solution of the gap equation corresponds to $\gamma =1$; 
the description provided by the latter might 
well be the closest approximation to reality. 
In fact, a more recent and complete analysis with deformation theory \cite{Poppitz_2}, moves
the estimate of the lower end -- most likely an upper bound of it -- of the conformal window 
for fundamental fermions to $\sim 4N$, close to the ladder prediction. The use of worldline 
formalism and large-$N$ \cite{Armoni_CW} leads to a similar estimate for fundamental 
fermions.
The rest of Fig.~\ref{fig:ConfPlot} will be treated in Section \ref{sec:slightly}.  

\begin{figure}
\begin{center}
\includegraphics[width=.55\textwidth]{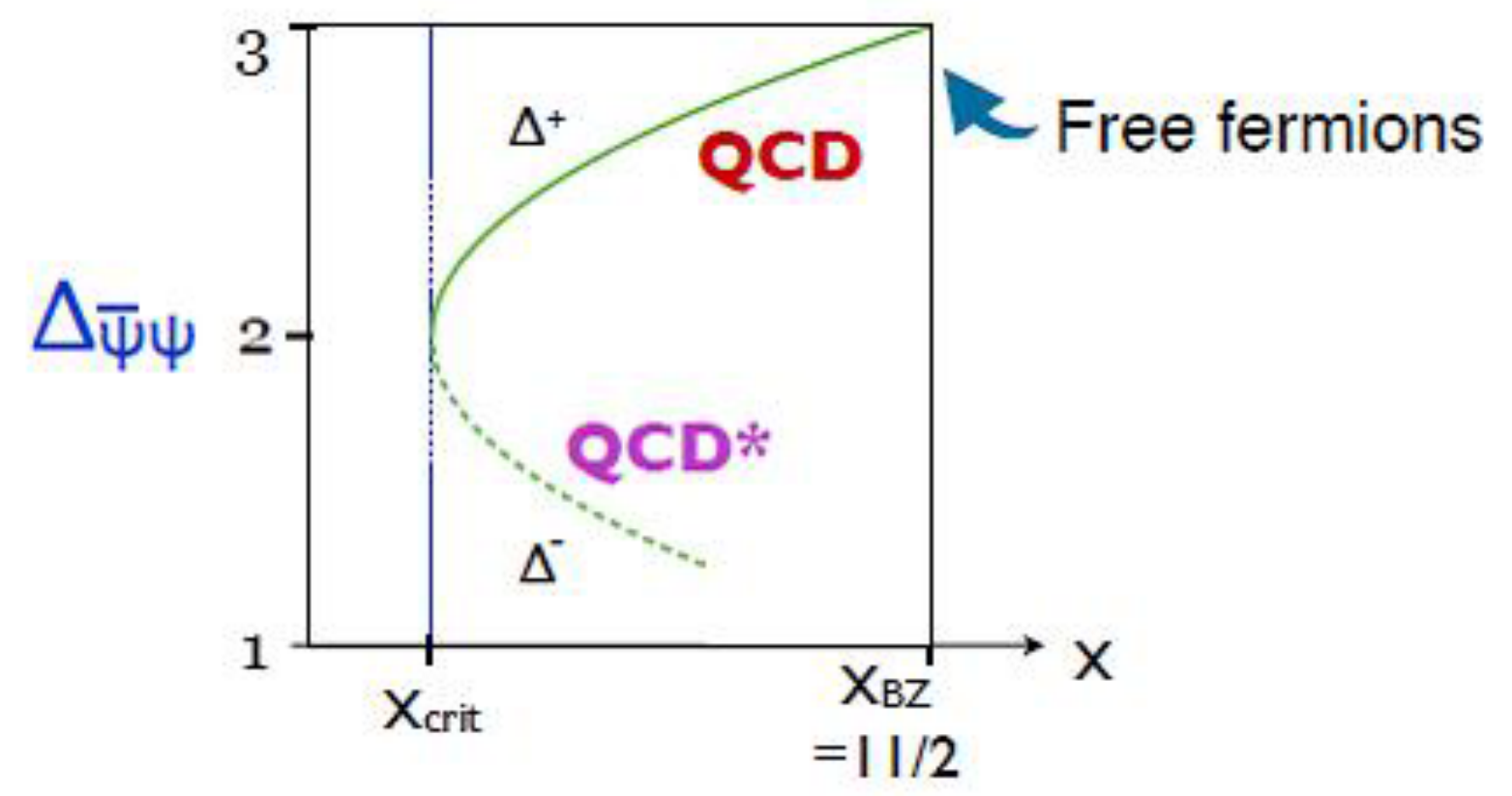}
\caption{Picture for the disappearance of conformality from \cite{Kaplan}, by varying $x=N_f/N_c$. The lines denote the dimension of the chiral condensate at the IR fixed point (upper)
 and the UV fixed point (lower). }
\label{fig:Kaplan_CFT}
\end{center}
\end{figure}

Already in \cite{BanksZaks} the coalescence of an infrared and ultraviolet fixed point 
was embedded in scenarios for conformality, although no conformal window or 
underlying mechanism for their merging was envisaged. 
Recently, the authors of \cite{Kaplan} have proposed a scenario for QCD at large number of 
flavors inspired by the AdS/CFT correspondence, where
the closure of the conformal window is due to the merging of the Banks-Zaks IR fixed point 
 and an UV fixed point, most plausibly appearing at strong coupling. 
The theories on the infrared 
 and ultraviolet  branch of Fig.~\ref{fig:Kaplan_CFT} are the boundary realization
 of a bulk scalar theory in AdS${}_5$. The dimension of the chiral condensate on the two 
branches is thus constrained by the profile of the bulk scalar field through AdS/CFT, 
so that the sum is fixed $\Delta_+ + \Delta_- =d=4$.
A relevant consequence is that the dimension of the chiral condensate on the IR branch 
ranges from 3 (the free limit) to 2, and never approaches the unitary limit $\Delta_+ =1$; 
this would create difficulties for those models of electroweak symmetry breaking 
demanding large -- greater than one -- anomalous dimensions. 
The merger mechanism, if in place, would distinguish QCD-like theories from their 
supersymmetric version, where the disappearance of the IRFP happens because 
$g_{IRFP}\to 0$ on one side of the conformal window where $N_f/N_c =3-\epsilon$, 
and $g_{IRFP}\to \infty$ on the other side where $N_f/N_c =3/2+\epsilon$.
 
Various elements hint at the possibility that the merger scenario would be
intimately connected with the scenario derived from the Schwinger-Dyson gap equation. 
In fact, known examples of merging (e.g. defect QFT) do produce a 
Berezinskii-Kosterlitz-Thouless (BKT) scaling of the chiral condensate \cite{Kaplan}, 
nothing but the Miransky scaling \cite{Miransky_conformal_1997} of a conformal phase 
transition.  
In addition \cite{Kaplan}, the point at which the anomalous dimension of 
the fermion mass operator becomes $\gamma =1$ is also the point at which four-fermi 
interactions 
$(\bar{\psi}\psi )^2$ become (marginally) relevant. This would suggest that QCD${}^*$, the 
dual theory of QCD on the UV branch, contains this relevant operator.  
As a consequence, the parameter space of the continuum effective theory at strong coupling 
is enlarged. 

A lattice study can in principle establish or exclude the existence of an ultraviolet 
stable fixed point at strong coupling. One example is the MBLL~\cite{MBLL} 
ultraviolet fixed point in quenched strongly coupled
QED, studied on the lattice in~\cite{Kogut_QED}. The first step of a lattice search 
would be to uncover 
the nature of the bulk transition observed in the vicinity of the lower end 
point of the conformal window -- this might be the case of \cite{Deuzeman_12} for $N_f=12$: 
a first order transition would exclude the occurrence of an UVFP in the continuum theory. 
One comment is in order about the presence of new relevant operators at strong coupling 
in the continuum effective lagrangian. There, the RG flow must be 
studied in a multi-dimensional parameter space. 
The lattice regularized theory, however, has only one adjustable bare parameter -- the gauge 
coupling. It is reasonable to expect that four-fermion like interactions will anyway be 
induced on the lattice by the gauge dynamics at sufficiently strong coupling;
 thus, we should not explicitly 
 account for the effective low energy realization of the continuum theory on the lattice. 
In other words, the physics of phase transitions extracted from the lattice regularized 
fundamental theory, should provide a complete description of the continuum strongly coupled 
theory.
\section{Slightly flavored theories: Other representations for fermions}
\label{sec:slightly}
The dependence of the two-loop beta function in eq.~(\ref{eq:beta}) upon the quadratic 
Casimir operator $C_2(R)$ and the 
trace $T(R)$ of the fermion representation $R$ shows that higher (than the fundamental) 
dimensional representations are more effective in i) loosing asymptotic freedom ii) 
acquiring an IRFP. In other words, they push the zero of the coefficients
$b_{0,1}$ to a lower number of flavors: for adjoint fermions in color 
$SU(3)$ -- see Table \ref{tab:R} -- asymptotic freedom is lost for $N_f^{AF}= 11/4=2.75$, 
while an IRFP -- 
a second zero of the beta function -- would appear for flavors larger than 
$N_f^* = 102/96\simeq 1.06$. 
Thus, it appears that slightly flavoring an $SU(N)$ gauge theory with fermions in higher 
dimensional representations would be sufficient for the emergence of conformality. 
The appearance of an infrared attractive fixed point 
 at low $N_f$ suggests new avenues for strongly-interacting theories beyond the 
Standard Model: the near appearance of a fixed point, in the proximity 
of a conformal window, with a beta function hover near 
the axis and a slowly varying gauge coupling, is an appealing feature for candidates of 
 walking technicolor theories. Phenomenology favors low $N_f$ and $N$. 
  
Turning to chiral dynamics, it is important to observe that the solution of the 
gap equation predicts, to leading order in the ladder approximation, a critical coupling 
$\alpha_c=\pi /(3C_2(R))$ for the emergence of a mass gap; such coupling decreases with 
increasing Casimirs. 
This argument shows that fermions in higher dimensional representations of the color group 
move the breaking of chiral symmetry to weaker couplings; 
short-distance forces alone become gradually more adequate to drive chiral symmetry breaking.
This is also in qualitative agreement with the Casimir scaling of the chiral symmetry 
breaking coupling $C_2(R) g_{ren}^2\simeq const$, observed in early quenched lattice 
studies~\cite{Kogut_1} for $SU(3)$ with sextet and octet fermions.

Lattice studies with dynamical fermions in varying representations 
should provide a better understanding of the interplay between  confinement and $\chi$SB
mechanisms. We must distinguish adjoint fermions, which do not feel confining forces at large
distances, from other representations that do feel confining forces: the latter will form 
bound states, a condition sufficient to induce $\chi$SB \cite{Casher}. In particular,
for $SU(N)$ with fundamental fermions, the 't Hooft 
anomaly matching conditions show that confinement implies 
chiral symmetry breaking for $N_f>2$~\cite{Frishman}. 
For adjoint fermions instead, the theory contains two natural scales, the string tension for 
heavy fundamental quarks and the $\chi$SB scale for light adjoint quarks.
A separation of scales was in fact observed \cite{KogutKarsch} for $SU(2)$ with 
two Dirac adjoint flavors, and interpreted as two thermal transitions with  
$T_{\mbox{deconf}}<T_{\mbox{chiral}}$.
\begin{figure}
\begin{center}
\includegraphics[width=.38\textwidth]{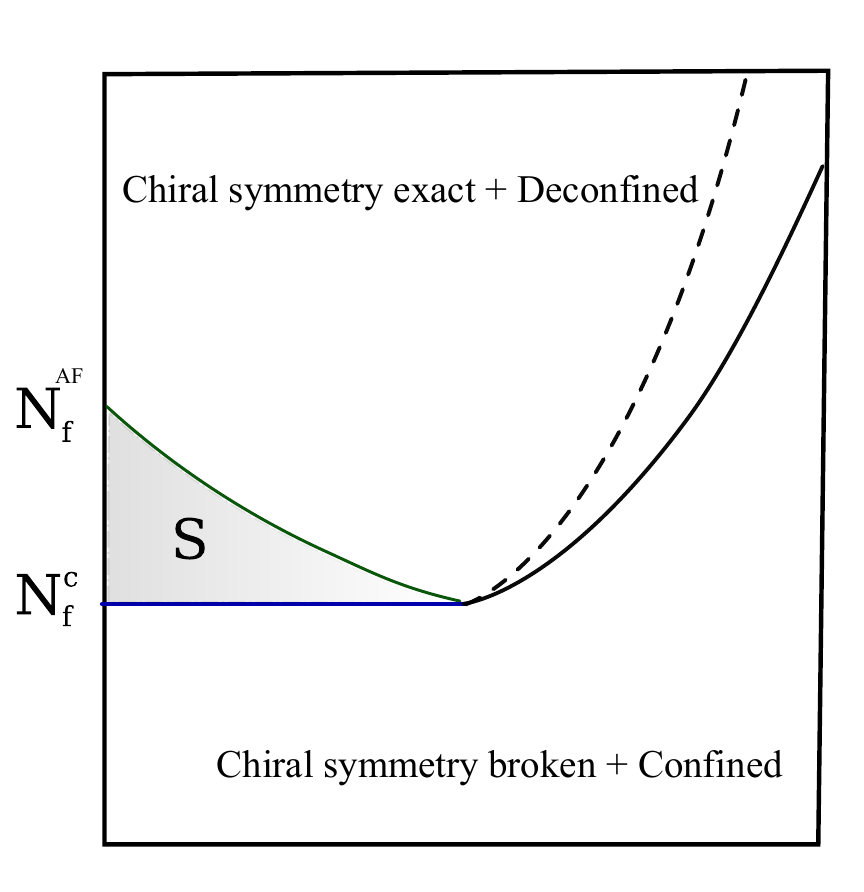}
\caption{ Speculations over the zero-temperature phase diagram for $SU(N)$ with 
adjoint fermions, assuming
that a conformal phase transition (horizontal blue line) occurs at $N_f^c$.
At strong coupling (r.h.s.) the ordering of the bulk  $\chi$SB transition (dashed) and 
confinement transition (solid) seems favored. 
At weak coupling (l.h.s.) the green line is the location of 
IRFPs. The shaded area  qualitatively indicates where a deconfinement transition may occur.
S stands for chirally symmetric.}
\label{fig:adjoint_plot}
\end{center}
\end{figure}

Based on these facts, we can attempt some speculations about the structure of the 
zero-temperature phase diagram for adjoint fermions and for other representations such as 
the two-index symmetric (2S) or antisymmetric (2A). For the latter two\footnote{In some cases representations may coincide, as for adjoint and 2S in $SU(2)$, or fundamental 
and 2A in $SU(3)$.} 
the formation of bound 
states will induce $\chi$SB, possibly leading to a similar phase diagram to fundamental 
fermions\footnote{A scale separation might still occur, with the $\chi$SB transition 
happening before the theory confines.}. 
Independently of the fermion representation, once $\chi$SB has occurred, fermions
acquire a dynamical mass and decouple at energies below this mass, leaving the pure gauge 
theory which confines.

If a conformal window arises through a conformal phase transition at a given $N_f^c$, as 
discussed for fundamental fermions, the phase diagram for adjoint flavors will be
different from the scenarios first depicted in \cite{BanksZaks}.
Instead, Fig.~\ref{fig:adjoint_plot} shows a possible scenario with a conformal phase 
transition. We have no theoretical prejudice about the relative order of the chiral and 
confinement transitions for adjoint fermions.           
However, at the lower end of the conformal 
window, call it $(N_f^c, g^c)$, the non perturbative beta function should turn negative, and 
confinement should follow. A plausible picture is that the confinement and chiral lines 
coincide at least at the critical point $(N_f^c,g^c)$. 
Away from it, the lines can a priori be disjoint, with the constraint that 
the theory is deconfined and chirally symmetric at the IRFP. 
The ordering of the bulk transitions on the right side of Fig.~\ref{fig:adjoint_plot} 
would be 
favored by \cite{KogutKarsch} and the occurrence of $\chi$SB at relatively weak coupling. 
At weaker coupling and for $g<g^c$ the theory is possibly confined for any flavor 
$N_f<N_f^c$.
The deconfinement transition might occur in the surroundings of the CPT line, 
or even coincide with the IRFP line\footnote{In this case the theory would be confining 
for all $N_f<N_f^{AF}$.}, as originally envisaged by the authors of \cite{BanksZaks}.
Their scenario of Fig.~3(b) in \cite{BanksZaks} would however be difficult to accommodate 
in a phase diagram 
where a conformal phase transition occurs.
 
The lower part of Fig.~\ref{fig:ConfPlot} reports on the 
conformal windows predicted for fermions in the adjoint and the two-index symmetric 
representations. Deformation theory \cite{Poppitz, Poppitz_2} and 
the gap equation produce similar estimates for the lower end point and
differ from the unitary bound prediction. Given that, according to \cite{Poppitz}, 
deformation theory  takes confinement into account, its 
agreement with the ladder prediction is suggestive of the fact that chiral dynamics alone, 
as previously argued, leads to a complete description of the emergence of conformality. 
For this reason, the initial disagreement of \cite{Poppitz} with the ladder prediction for
fundamental fermions was rather surprising, and recently resolved in 
\cite{Poppitz_2}\footnote{Notice that, differently from the gap equation ladder-approach 
\cite{Appelquist_ladder}, the formalism of \cite{Poppitz,Poppitz_2} can still be applied to
 chiral gauge theories which exhibit confinement without chiral symmetry breaking.}.
Additional estimates with the worldline formalism for various gauge groups and matter 
representations can be found in \cite{Armoni_CW}.
 
Lattice studies are currently being carried out in two specific cases, 
both candidates for technicolor theories: $SU(3)$ with two Dirac flavors in the 
two-index symmetric (2S) representation of the color group~\cite{Degrand}, and in the quenched approximation \cite{Kuti_2S}, 
and $SU(2)$ with two Dirac flavors in the adjoint (Adj) 
representation~\cite{Adjoint_Lat, Lucini,Hietanen_RG}.

The study in \cite{Degrand} favors the presence of an 
IRFP for $SU(3)$ with two sextet (2S) fermions, by using the discrete beta function and 
studying the confinement phase transition in the lattice 
parameter space with varying volume. The degeneracies of the screening masses and behavior 
of the pseudoscalar decay constant indicate that the deconfined phase is also a phase 
in which chiral symmetry is restored.
Note that the ladder solution of the gap equation and deformation theory
predict a lower end of the conformal window at $N_f^c \simeq 2.5$ and $N_f^c \simeq 2.4$, 
respectively, while asymptotic freedom is lost at $N_f^{AF}=3$. 
Hence, a simulation at $N_f =2$ is likely to be in the vicinity 
of the conformal window, rendering a lattice study a delicate task. 
For $N_f=1$ and in the large-$N$ limit the theory is non perturbatively equivalent to
 supersymmetric Yang-Mills \cite{Armoni}, and thus confines.

It has been proposed that four-dimensional QCD with adjoint fermions exhibits volume 
independence in the large-$N$ limit \cite{EguchiKawai}. This would allow for 
{\em reduction}, i.e. the possibility to determine the properties of the theory from just 
a single-site lattice calculation. 
Evidence for reduction is found in \cite{Sharpe} with one flavor of Dirac adjoint (Wilson) 
fermions. Another study \cite{Narayanan_LAT09}, using overlap as opposed to Wilson fermions,
has recently confirmed the evidence for reduction.

Analytical predictions for the lower end of the conformal window suggest, as shown in 
Fig.~\ref{fig:ConfPlot}, 
that the $SU(2)$ theory with two Dirac fermions in the adjoint representation is a likely 
candidate for a near-conformal or walking scenario. 
Lattice calculations of the spectrum of the theory \cite{Adjoint_Lat} and the 
renormalization group flow
 \cite{Hietanen_RG} possibly hint at a conformal or near-conformal behavior, though still 
requiring an improved control of systematic uncertainties.  
Recently, the authors of \cite{Lucini} have outlined a strategy based on the combination of 
gluonic and mesonic observables 
in order to discriminate between a confining and conformal scenario. 

\section{Some concluding remarks}
\label{sec:conc}

The last two years have seen a rapidly growing interest in the yet unexplored aspects of non 
abelian gauge theories. The list of names, contributing to this enterprise, is getting 
longer.
We interrogate over the emergence, or loss, of conformal symmetry by varying the flavor
 content of the theory; what is the interplay of chiral dynamics and confining forces? 
What residual analogy does survive, when going 
from the supersymmetric to the non supersymmetric theory? 
Elucidating the way conformal symmetry or its remnants drive particle dynamics will 
contribute to clarifying the possible connection of field theory to string theory that
the AdS/CFT correspondence seems to imply.  
Lattice simulations are the natural place where to look for a non-perturbative answer 
and complement analytical predictions and conjectures. 

One main motivation, genuinely theoretical, drives these lattice studies: we aim 
at a better understanding of the phase diagram of gauge theories. This is a reason enough
to pursue this goal. Such an understanding is not only important per se, but it is likely to 
improve our way and tools for describing the critical behavior and the breaking of 
symmetries in field theory, where relevant examples go from particle physics to condensed 
matter systems.  
All the subtleties of field theory, on and off the lattice, come into play in this 
entertaining game.

A connection to the physics at the Large Hadron Collider (LHC) is also a must, in these 
times. Needless to say that conformal symmetry might play an important role beyond the 
electroweak symmetry breaking (EWSB) scale, and as such it is potentially relevant for 
physics searches at the LHC.
Quark-gluon plasma physics at Alice will teach about deviations from conformality, and  
LHC will say about the mechanism of EWSB: is it induced by 
a strongly interacting dynamics, of which walking technicolor is an example, and what gauge 
groups do realize the phenomenologically allowed symmetry breaking pattern?
More in general, and without invoking technicolor, the unification of forces might
still be realized in a scenario where gauge groups attain a quasi-conformal dynamics, with 
or without supersymmetry, for a given interval of energies.
The impact on phenomenology, with the LHC activities just starting, added to a theoretically 
driven curiosity, make it the right time to pursue these studies.

\acknowledgments 

I thank especially my collaborators A. Deuzeman and M. P. Lombardo for the 
enjoyable and fruitful work together. I also thank A. Armoni, J. Braun, J. Kuti, 
B. Svetitsky and M. \"Unsal for discussions and correspondence.


\begin{thebibliography}{99}

\bibitem{Tumbling}
S. Raby, S. Dimopoulos, L. Susskind, Nucl.\ Phys.\  B {\bf 169} (1980) 373.
\bibitem{Walking}
A. Cohen, H. Georgi, Nucl.\ Phys.\ B {\bf 314} (1989) 7, and refs therein. 
\bibitem{AsymFree} 
D.J. Gross, F. Wilczek,  Phys.\ Rev.\ Lett.\  {\bf 30} (1973) 1343; 
H.D. Politzer,  Phys.\ Rev.\ Lett.\  {\bf 30} (1973) 1346.  
\bibitem{BraunGies}
J. Braun, H. Gies, JHEP 024 (2006); J. Braun, H. Gies, arXiv:0912.4168.
\bibitem{Caswell}
   W.~E.~Caswell,
   Phys.\ Rev.\ Lett.\  {\bf 33} (1974) 244.
\bibitem{BanksZaks}
  T.~Banks and A.~Zaks,
  Nucl.\ Phys.\  B {\bf 196} (1982) 189.
\bibitem{Kaplan}
D.B. Kaplan, J.-W. Lee, D.T. Son, M.A. Stephanov, arXiv:0905.4752.
\bibitem{Miransky_conformal_1997}
V.A.~Miransky, K.~Yamawaki, Phys.\ Rev.\ {\bf D55} (1997) 5051; Erratum-ibid.D56 (1997) 
3768. 
\bibitem{Appelquist_1996}
T. Appelquist, J. Terning, L.C.R. Wijewardhana, Phys.\ Rev.\ Lett.\  {\bf 77} (1996) 1214;
T. Appelquist, A. Ratnaweera, J. Terning, L.C.R. Wijewardhana,  Phys.\ Rev.\ {\bf D58} 
(1998) 105017.
\bibitem{Appelquist_ladder}
T.  Appelquist, K. Lane, U. Mahanta, Phys.\ Rev.\ Lett.\  {\bf 61} (1988) 1553;
T.  Appelquist, A. G. Cohen, M.  Schmaltz,  Phys.\ Rev.\ {\bf D60} (1999) 
045003.
\bibitem{Appelquist_RG}
  T.~Appelquist, G.~T.~Fleming and E.~T.~Neil,
  Phys.\ Rev.\ Lett.\  {\bf 100} (2008) 171607; Phys.\ Rev.\ {\bf D79} (2009) 076010. 
\bibitem{Hasenfratz}
A. Hasenfratz, arXiv:0911.0646; arXiv:0907.0919. 
\bibitem{Deuzeman_12}
A.~Deuzeman, M.~P.~Lombardo, E.~Pallante, arXiv:0904.4662.
\bibitem{Heller}
  P.~H.~Damgaard, U.~M.~Heller, A.~Krasnitz and P.~Olesen,
  Phys.\ Lett.\  B {\bf 400} (1997) 169
\bibitem{Peskin}
M.E. Peskin, Nucl.\ Phys.\ {\bf B175} (1980) 197.
\bibitem{Svetitsky} 
P.H. Damgaard, U.M. Heller, R. Niclasen, B. Svetitsky,  Nucl. Phys. {\bf B633} (2002) 97.
\bibitem{Kuti_12}
Z. Fodor, K. Holland, J. Kuti, D. Nogradi, C. Schroeder, arXiv:0911.2463;
Phys.\ Lett.\  B {\bf 681} (2009) 353.
\bibitem{Deuzeman_8}
A.~Deuzeman, M.~P.~Lombardo, E.~Pallante,  Phys.\ Lett.\  B {\bf 670} (2008) 41;
X.-Y. Jin, R.D. Mawhinney, PoS(LAT2008){\bf 59}.
\bibitem{Mawhinney_12}
X.-Y. Jin, R.D. Mawhinney, PoS(LAT2009){\bf 49}.
\bibitem{Progress}
A.~Deuzeman, M.~P.~Lombardo, E.~Pallante, in progress.
\bibitem{Sannino_AD}
F. Sannino, Phys.\ Rev.\ {\bf D80} (2009) 017901.
\bibitem{Seiberg}
N. Seiberg,  Nucl. Phys. {\bf B435} (1995) 129.
%
\bibitem{NSVZ}
V.A. Novikov, M.A. Shifman, A.I. Vainshtein, V.I. Zakharov, Nucl. Phys. {\bf B229} (1983) 
381;   M.A. Shifman, A.I. Vainshtein, Nucl. Phys. {\bf B277} (1986) 456.
\bibitem{Armoni}
A. Armoni, M. Shifman, G. Veneziano, Phys.\ Rev.\ Lett.\  {\bf 91} (2003) 191601; 
Nucl.\ Phys.\ {\bf B667} (2003) 170.  
\bibitem{Bochicchio}
M. Bochicchio, JHEP 0905:116, 2009.
%
\bibitem{Sannino_Susy}
T.A. Ryttov, F. Sannino, Phys.\ Rev.\ {\bf D78} (2008) 065001.
%
\bibitem{Poppitz}
E. Poppitz, M. \"Unsal, arXiv:0906.5156.
%
\bibitem{Poppitz_2}
E. Poppitz, M. \"Unsal, arXiv:0910.1245.
\bibitem{Armoni_CW}
A. Armoni,  Nucl. Phys. {\bf B826} (2010) 328.
\bibitem{MBLL}
C.N. Leung, S.T. Love, W. Bardeen, Nucl. Phys. {\bf B273} (1986) 649.
\bibitem{Kogut_QED}
A. Kocic, S. Hands, J.B. Kogut, E. Dagotto, Nucl. Phys. {\bf B347} (1990) 217.
\bibitem{Kogut_1} 
J.B. Kogut, J. Shigemitsu, D.K. Sinclair,  Phys.\ Lett.\  B {\bf 145} (1984) 239.
\bibitem{Casher}
A. Casher,  Phys.\ Lett.\  B {\bf 83} (1979) 395;
T. Banks, A. Casher, Nucl.\ Phys.\ {\bf B169} (1980) 103;
S. Coleman, E. Witten,  Phys.\ Rev.\ Lett.\  {\bf 45} (1980) 100.
\bibitem{Frishman} 
G. 't Hooft, 1979 Cargese School;  
Y. Frishman, A. Schwimmer, T. Banks, S. Yankielowicz, Nucl.\ Phys.\ {\bf B 177} (1981) 157.
\bibitem{KogutKarsch}
J.B. Kogut, J. Polonyi, H.W. Wyld, Phys.\ Rev.\ Lett.\  {\bf 54} (1985) 1980;
F. Karsch, L\"utgemeier,  Nucl. Phys. {\bf B550} (1999) 449; Nucl.\ Phys.\ B (Proc. Suppl.) 
73 (1999) 446.
\bibitem{Degrand}
Y. Shamir, B. Svetitsky and T. DeGrand, Phys.\ Rev.\ {\bf D78} (2008) 031502;
T. DeGrand, Y. Shamir and B. Svetitsky, Phys.\ Rev.\ {\bf D79} (2009) 034501.
\bibitem{Kuti_2S}
Z. Fodor, K. Holland, J. Kuti, D. Nogradi, C. Schroeder, JHEP 0911:103, 2009; JHEP 0908:084, 2009.
\bibitem{Adjoint_Lat}
S. Catterall, F. Sannino, Phys.\ Rev.\ {\bf D76} (2007) 034504; S. Catterall, J. Giedt, F. Sannino, J. Schneible, JHEP {\bf 11}, 009 (2008); 
L. Del Debbio, A. Patella, C. Pica, arXiv:0805.2058; arXiv:0812.0570; 
A.J. Hietanen, J. Rantaharju, K. Rummukainen, K. Tuominen, arXiv:0812.1467.
\bibitem{Lucini}
L. Del Debbio, B. Lucini, A. Patella, C. Pica, A. Rago, Phys.\ Rev.\ {\bf D80} (2009) 074507.
\bibitem{Hietanen_RG}
A.J. Hietanen, K. Rummukainen, K. Tuominen, arXiv:0904.0864.
\bibitem{EguchiKawai}
T. Eguchi, H. Kawai,  Phys.\ Rev.\ Lett.\  {\bf 48} (1982) 1063;
P. Kovtun, M. \"Unsal, L.G. Yaffe, JHEP 0706:019, 2007.
\bibitem{Sharpe}
B. Bringoltz, S.R. Sharpe, Phys.\ Rev.\ {\bf D80} (2009) 065031.
\bibitem{Narayanan_LAT09}
A. Hietanen, R. Narayanan, arXiv:0911.2449.
 \end{thebibliography}
 \end{document}